\documentclass{EJAM}

\usepackage{epstopdf}
\usepackage{upgreek}
\usepackage{graphicx} 
\usepackage{hyperref}
\usepackage{diagbox}
\usepackage[dvipsnames]{xcolor}
\usepackage{amsmath}
\usepackage{amsfonts}
\usepackage{subcaption}
\usepackage{bm}
\usepackage{caption}
\usepackage{algorithm}
\usepackage{diagbox}
\usepackage{booktabs}
\usepackage{multirow}
\usepackage{algpseudocode}

\msctttrue
\allowdisplaybreaks

\usepackage[version=4]{mhchem}
\usepackage{amsthm}
\graphicspath{ {./Plots/} }
\hypersetup{pdfborder=0 0 0}
\usepackage{multirow}
\usepackage{booktabs}
\usepackage{makecell}
\usepackage{array}
\usepackage{tabularx}
\usepackage{ragged2e}
\usepackage{makecell}
\newcolumntype{L}[1]{>{\RaggedRight\arraybackslash}p{#1}}
\newcolumntype{C}[1]{>{\centering\arraybackslash}p{#1}}
\newcolumntype{Y}{>{\RaggedRight\arraybackslash}X}
\newtheorem{hypothesis}{Hypothesis}
\newsavebox{\astrutbox}
\sbox{\astrutbox}{\rule[-5pt]{0pt}{20pt}}

\begin{document}

\jnlDoiYr{2026}
\doival{XX.XXXX/xxxxx}

\lefttitle{Bansal N. {\it et al.}}
\righttitle{European Journal of Applied Mathematics}

\papertitle{RESEARCH ARTICLE}

\title{Extending indoor airborne infection risk modelling to 3D, with local surface deposition}
\author{Neha Bansal}
\author{Katerina Kaouri}
\author{Thomas E. Woolley}

\affil{School of Computational and Mathematical Sciences, Cardiff University, Cardiff, United Kingdom \\ \textbf{Corresponding author:} Neha Bansal; Email: \href{mailto:bansaln3@cardiff.ac.uk}{bansaln3@cardiff.ac.uk}}

\received{XX XX XXXX}\revised{XX XX XXXX}\accepted{XX XX XXXX}

\begin{abstract}

Mathematical models of indoor disease transmission often rely on simplifying assumptions to reduce computational cost. In particular, in the Viral Infection Risk Indoor Simulator (VIRIS) \cite{xue_viris_2024}, which combines architectural layout, people movement and airborne viral transmission, aerosol concentrations are assumed uniform across room height, allowing a two-dimensional (2D) formulation. Also, particle deposition on surfaces is represented by a first-order removal term acting throughout the computational domain. Here, we extend VIRIS to the VIRIS-3D-L framework to investigate how these two assumptions affect the predicted infection risk. In VIRIS-3D-L, we solve the reaction-diffusion equation in three dimensions (3D), and aerosol deposition is represented by local surface removal through a Robin boundary condition. We compare VIRIS with VIRIS-3D-L across three indoor settings of increasing architectural and people-movement complexity: a courtroom, a care home, and a supermarket. For the courtroom, care home, and supermarket, respectively, at the end of the simulation, the original 2D formulation with global removal predicts average infection risks across individuals of 44\%, 78\%, and 25\%, while VIRIS-3D-L predicts risks of 51\%, 85\%, and 28\%. Thus, VIRIS significantly underestimates infection risk in all three settings. The predictions of VIRIS-3D-L also show agreement with a documented courtroom outbreak that was used to validate VIRIS. Although computation time increases by approximately a factor of four on average, simulations remain feasible on a standard laptop, taking between eight minutes and just under 1.5 hours. These findings suggest that VIRIS-3D-L offers a more physically realistic framework for assessing indoor airborne infection risk in indoor spaces without prohibitive computational costs.
\end{abstract}


\begin{msc}
{\it Primary} - 92D30, 92-10, 76-10;
{\it Secondary} - 35E15 
\end{msc}

\maketitle

\section{Introduction}\label{sec:intro}
The Viral Infection Risk Indoor Simulator (VIRIS) \cite{xue_viris_2024} is a holistic modelling framework powered by Topologicpy \cite{jabi2024topologicpy} that combines architectural layout, people movement, and airborne viral transmission. VIRIS can be customised to represent a variety of indoor settings, people behaviour, and infectious diseases, in a computationally efficient manner. For example, disease spread for tens of individuals moving in an indoor space over the course of the day can be simulated in seconds on a standard laptop. 

The airborne transmission model in VIRIS has two components \cite{xue_viris_2024}. First, a reaction-diffusion equation, adapted from \cite{lau_predicting_2022}, governs the spatio-temporal concentration of infectious aerosols. A turbulent eddy diffusivity \cite{Foat_Drodge_Nally_Parker_2020} is used in the reaction-diffusion equation to model the effect of local turbulent air mixing. Second, infection risk is determined following \cite{lau_predicting_2022}, who developed an extension of the Wells-Riley ansatz.

In VIRIS, infectious aerosols are assumed to be uniformly distributed throughout the height of the building; therefore, airborne transmission is modelled in a two-dimensional (2D) domain. However, as aerosols diffuse in three dimensions, there are vertical concentration gradients and higher concentrations of infectious aerosols in the breathing zone. Consequently, the two-dimensional formulation leads to an underestimation of aerosol concentrations at an individual's mouth level and cannot represent variations in aerosol transport due to differences in people's heights and furniture heights. These limitations affect the predictions of transmission dynamics and infection risk. Additionally, the 2D formulation renders VIRIS unsuitable for studying multi-storey buildings. Extending VIRIS to a three-dimensional (3D) domain addresses these limitations; this is one of the extensions we present in this work.

Furthermore, in VIRIS, surface deposition of aerosols is represented as a first-order global removal term in the reaction-diffusion equation, following several airborne transmission models \cite{lau_predicting_2022,miller2021transmission,gammaitoni1997using}. As aerosols deposit on all surfaces such as walls, ceilings, floors, furniture, or other solid objects, the global removal assumption leads to an underestimation of aerosol concentrations in the breathing zone, and consequently to an underestimation of the infection risk. In this study, we remove the global removal assumption and present a modified model with a more realistic, local-removal formulation.

Since the 2D domain and global removal assumptions result in lower concentrations of infectious aerosols, they lead to an underestimation of infection risk. This, in turn, could lead to insufficient disease control measures. We, therefore, modify the reaction-diffusion equation in VIRIS for solving in a 3D domain and replace the global surface deposition term with a Robin boundary condition that represents local removal at surfaces; we call this new modelling framework VIRIS-3D-L. Additionally, VIRIS-3D-L enables the study of multi-storey buildings, allowing vertical heterogeneity and connections between floors, which are important factors to account for. 

The paper is organised as follows. In Section~\ref{sec:methods}, we describe the VIRIS-3D-L framework. In Sections~\ref{sec:plots_2d_3d_3dl} and~\ref{sec:comp_time}, respectively, we compare the infection risk predictions and computation times of VIRIS and VIRIS-3D-L for three single-storey settings with increasing architectural and people movement complexity: a courtroom, a care home, and a supermarket. Finally, in Section~\ref{sec:conclusion}, we present our conclusions and directions for future work.

\section{Mathematical model}\label{sec:methods}
In this section, we first briefly describe the VIRIS modelling framework \cite{xue_viris_2024}. Next, we describe the new framework we developed, VIRIS-3D-L. 

\subsection{VIRIS: Airborne transmission modelling in a 2D domain}\label{sec:viris_og}
In the VIRIS framework \cite{xue_viris_2024}, people’s behaviour is modelled using an agent-based model. An infectious person exhales viral aerosols, and a susceptible person becomes infected by inhaling a certain amount of infectious aerosols. People are assumed to be points and not full-body solid objects. Aerosol transport is assumed to be by diffusion, driven by spatial concentration gradients. Advection and drag forces are neglected, as airflow-driven aerosol transport can be neglected in poorly ventilated settings \cite{xue_viris_2024,bhagat2024building,bhagat2020effects}. The impact of local turbulent air mixing on diffusion is modelled using a turbulent eddy diffusivity \cite{Foat_Drodge_Nally_Parker_2020,shao2017turbulent}, $$D = 0.8 \lambda V^{2/3},$$ where $\lambda$ is the air exchange rate and $V$ is the volume of the space. Airborne transmission dynamics are simulated in a 2D domain by assuming uniform aerosol concentration across the height. Infectious aerosols are assumed to be removed due to three factors: virus deactivation, surface deposition, and ventilation. A first-order, global removal term is used to depict each of these three removal processes.

\subsubsection{Airborne transmission modelling in a 2D domain}\label{sec:viris_eqs}
Let $\Omega \in \mathrm{R}^2$ be a 2D domain with $n$ individuals, of which $m$ are infectious. Let $C(\mathbf{x},t)$ be the concentration of infectious aerosols at location $\mathbf{x} \in \Omega$ and at time $t$. Then the rate of change in $C$ is given by:
\begin{align} \label{eq:rd_viris_2d}
    \frac{\partial C(\mathbf{x},t)}{\partial t} = D\nabla^2C -(\kappa+R)C+ S, \forall \mathbf{x} \in \Omega, \text{ and } t \in [0,T],
\end{align}
where $\kappa = \beta + \lambda + \gamma$ is the first-order global removal term, where $\beta$ is the viral deactivation rate \cite{dabisch2021influence,aganovic2021estimating}, $\lambda$ is the air exchange rate and $\gamma$ is the surface deposition rate. $R$ is the sink term from aerosol inhalation by all individuals, and $S$ is the source term from aerosol exhalation by infectious individuals.

The amount of infectious aerosols in the domain due to $m$ infectious individuals, the source term, $S(\mathbf{x},t)$, is defined as:
\begin{align}\label{eq:source_term}
    S(\mathbf{x},t) &= \sum_{i=1}^{m} \frac{(1-\eta_i)}{h}A_i\mu \delta(\mathbf{x}-\mathbf{x_i}(t)),
\end{align}
where $\eta_i$ is the mask efficiency of the $i^{\text{th}}$ individual, $h$ is the room height, $A_i$ is the aerosol emission rate of the $i^{\text{th}}$ individual, $\mu$ is the fraction of virus-laden aerosols, and $\delta$ is the Dirac delta function representing a point source. 

The sink due to inhalation by all $n$ individuals in the domain, $R(\mathbf{x},t)$, is defined as:
\begin{align}\label{eq:sink_term}
    R(\mathbf{x},t) &= \sum_{i=1}^{n} (1-\eta_i)\rho_i \delta(\mathbf{x}-\mathbf{x_i}(t)),
\end{align}
where $\rho_i$ is the breathing rate of the $i^{\text{th}}$ individual.

A Neumann boundary condition is used, assuming perfect reflection of the aerosols off the walls, 
\begin{align}\label{eq:neumann}
    \frac{\partial C(\mathbf{x},t)}{\partial n}  &= 0, \quad \mathbf{x} \in \partial\Omega \text{ and } t \in [0,T],
\end{align}
where $n$ is the outward normal direction to $\Omega$, and $\partial\Omega$ is the boundary of the domain. The initial concentration of infectious aerosol is assumed to be $0$, i.e.,
\begin{align}\label{eq:init_condt}
    C(\mathbf{x},0) & =0,\quad  \forall \mathbf{x} \in \Omega.
\end{align}

\subsubsection{Infection risk}\label{sec:viris_inf_risk}
We use the spatio-temporal concentration of infectious aerosols, obtained by solving Eq. \eqref{eq:rd_viris_2d}, to evaluate the infection risk, following \cite{xue_viris_2024} and \cite{lau_predicting_2022}, who extended the Wells-Riley ansatz to a spatio-temporal ansatz. The infection risk for a susceptible individual with mask efficiency $\eta$ over the exposure period from $0$ to $t$ is:
\begin{align}\label{eq:inf_risk_3D}
    \mathcal{P}(t) =  1- \exp\Big( -I \int_{0}^t (1-\eta) \rho C(\mathbf{x}({\tau}),\tau) d\tau\Big),
\end{align}
where $\mathbf{x}({\tau})$ is the individual's location at time $\tau$, and
\begin{align*}
I = -\frac{\ln (0.5)}{d_m},  
\end{align*}
where $d_m$ is the median infectious dose of infectious aerosols \cite{qiu2023site}. 

\subsection{VIRIS-3D-L: Airborne transmission modelling in a 3D domain with local removal due to  surface deposition}\label{sec:viris_3d_l}
In VIRIS, the global removal assumption is reasonable because transmission is modelled in a 2D domain; removing aerosols uniformly across the entire domain accounts for deposition on the ceiling, floor, walls, and other furniture surfaces. However, in a 3D domain, concentrations vary with height, allowing us to distinguish aerosol concentrations at surfaces from those in the remaining domain. Therefore, global removal is not appropriate; we need to remove only the infectious aerosols that deposit on surfaces.

In the modified modelling framework VIRIS-3D-L, we modify the modelling assumptions of VIRIS (Section \ref{sec:viris_og}) as follows.

First, we represent each person as a point source or sink at a height of 1.8m, and thus, model the $i^{\text{th}}$ individual at time $t$ as a point source (sink) at $(x_i(t), y_i(t),1.8\text{m})$. (Although we could use a custom height for each person as input in VIRIS, we take all heights to be equal for simplicity.) 

Second, we solve the reaction-diffusion equation for airborne transmission, given by Eqs. \eqref{eq:rd_viris_2d}-\eqref{eq:sink_term}, in three dimensions, $\Omega \in \mathrm{R}^3$. 

Since we represent local removal due to surface deposition using a Robin boundary condition \cite{evans2010pde}, we remove the first-order global removal term, $\gamma$, from Eq. \eqref{eq:rd_viris_2d}. We assume all surfaces have the same removal rate. Thus, the equation is
\begin{align} \label{eq:rd_viris_3d}
    \frac{\partial C(\mathbf{x},t)}{\partial t} = D\nabla^2C -(\kappa+R)C+ S, \forall \mathbf{x} \in \Omega, \text{ and } t \in [0,T],
\end{align}
where $\kappa = \beta + \lambda $. We also remove the $1/h$ term in Eq. \eqref{eq:source_term}, since we no longer assume uniform concentration across the room height.

The Neumann boundary condition (Eq.~\eqref{eq:neumann})
is replaced by the Robin boundary condition
\begin{equation}
    D\frac{\partial C(\mathbf{x},t)}{\partial n}  = -v_w F C, \quad \mathbf{x}\in \partial\Omega \text{ and } t \in [0,T],
    \label{eq:robin}
\end{equation}
where $n$ is the outward unit normal to the domain,
$v_w$ is the deposition velocity, and $F$ is a dimensionless
scaling factor chosen such that
\begin{equation}
    \frac{v_w F}{D}
    = 1.1 \times 10^{-4}\,\mathrm{m}^{-1}.
    \label{eq:deposition_coefficient}
\end{equation}
This coefficient $v_w F/D$ is held fixed across
all simulated settings. The initial condition
(Eq.~\eqref{eq:init_condt}) and the infection-risk formulation
(Eq.~\eqref{eq:inf_risk_3D}) remain unchanged.


As an intermediate methodological step, we also developed and simulated VIRIS-3D, a framework in which the reaction-diffusion equation (\eqref{eq:rd_viris_2d}) is solved in a 3D domain while retaining the first-order global removal term for surface deposition as in VIRIS. The factor \(1/h\) is removed from the source term in Eq. \eqref{eq:source_term}, since aerosol concentration is no longer assumed uniform across room height. We briefly present results from VIRIS-3D in Section \ref{sec:results} to help distinguish the effects of extending the model to 3D from those of replacing global removal with local surface deposition.

\subsection{Numerical implementation and parameter values: comparison of VIRIS and VIRIS-3D-L}\label{sec:sim_param_process}
We generate a mesh of the indoor space geometry using the GMSH Python library \cite{geuzaine2009gmsh}. To compare VIRIS and VIRIS-3D-L, we generate two meshes with an element size of $0.2$m: one for a 2D domain with triangular elements and another for a 3D domain with tetrahedral elements.

In VIRIS \cite{xue_viris_2024}, the Scikit-FEM Python library  \cite{gustafsson2020scikit} is used to solve the reaction-diffusion equation \eqref{eq:rd_viris_2d} along with its built-in point-source function to represent the delta functions in Eq. \eqref{eq:source_term} and Eq. \eqref{eq:sink_term}. In this paper, for both frameworks, we use the FEniCS Python library \cite{baratta2023dolfinx} to solve the reaction-diffusion equations. Further, we approximate the delta function using an isotropic Gaussian distribution \cite{pretty2026effect,squires2001practical}, with a standard deviation of $ 0.2$m, aligned with the mesh element sizes.

We simulate airborne transmission in three settings: a courtroom, a care home, and a supermarket, as in \cite{xue_viris_2024}. The details of the architectural plans and people schedules are provided in Appendix \ref{layouts_3} and \cite{xue_viris_2024}. We use the SARS-CoV-2 disease parameters listed in Table \ref{tab:simulation_params} to estimate infection risk. For all three settings, we assume poor ventilation conditions. 

The courtroom simulations are informed by the superspreader outbreak data \cite{vernez_probable_2021}. We represent the infectious individual in the courtroom as a superspreader and those in the care home and supermarket as non-superspreaders by assigning different values to the fraction of virus-laden aerosols, $\mu$ (Table \ref{tab:simulation_params}). For the courtroom, we fix $\mu=0.325$, the assumed superspreader value reported for the outbreak \cite{vernez_probable_2021}, and calibrate the median infection dose $d_m$ to match the data on the number of infected individuals \cite{vernez_probable_2021}. At $\mu = 0.325$, $d_m=300$ maximises the likelihood of reproducing the outbreak data for both VIRIS and VIRIS-3D-L. Appendix \ref{app:courtroom_mu_dm} details the parameter sweep and likelihood calculation. 

We consider three activity types: resting, talking, and talking loudly. The breathing rate ($\rho$) and aerosol emission rate ($A$) are listed in Table \ref{tab:simulation_params}. All simulations use a six-second time-step, taken to represent one breathing cycle. The simulated durations are three hours for the courtroom, 15 hours for the care home, and five hours for the supermarket. Simulations are run on a standard laptop. The code is available on \href{https://github.com/nehabansal26/VIRIS_extensions_3D_and_local_removal}{GitHub}.

\begin{table}[htbp]
\centering
\small
\caption{Simulation parameters for SARS-CoV-2.}
\label{tab:simulation_params}

\renewcommand{\arraystretch}{1.18}
\setlength{\tabcolsep}{4pt}

\begin{tabularx}{\linewidth}{
    L{5.0cm}
    C{1.2cm}
    L{1.9cm}
    C{2.0cm}
    Y
}
\toprule
\textbf{Parameter} & \textbf{Symbol} & \textbf{Setting} & \textbf{Value} & \textbf{References} \\
\midrule
Average aerosol diameter ($\mu\mathrm{m}$)
& $d$
& All settings
& $5$
& \cite{lau_predicting_2022}
\\
\hline
Virus deactivation rate ($\mathrm{s}^{-1}$)
& $\beta$
& All settings
& $1.7\times10^{-4}$
& \cite{van2020aerosol,aganovic2021estimating}
\\
\hline
Surface deposition rate ($\mathrm{s}^{-1}$)
& $\gamma$
& All settings
& $1.1\times10^{-4}$
& \cite{de2022modeling}
\\
\hline
Deposition velocity ($\mathrm{m/s}$)
& $v_w$
& All settings
& $7.6\times10^{-4}$
& \cite{wang2021airborne}
\\
\hline
\multirow{3}{4.0cm}{Air change rate ($\mathrm{s}^{-1}$)}
& \multirow{3}{0.8cm}{\centering $\lambda$}
& Courtroom
& $6.4\times10^{-5}$
& \multirow{3}{\linewidth}{\cite{
    vernez_probable_2021,
    lau_predicting_2022,
    ASHRAE62.1_2022_ab}}
\\
& & Care home
& $3.3\times10^{-5}$
& \\
& & Supermarket
& $3.3\times10^{-5}$
& \\
\addlinespace
\hline
\multirow{3}{4.0cm}{Fraction of virus-laden aerosol}
& \multirow{3}{0.8cm}{\centering $\mu$}
& Courtroom
& $0.325$
& \multirow{3}{\linewidth}{\cite{vernez_probable_2021,
        lau_predicting_2022}
    }
\\
& & Care home
& $0.065$
& \\
& & Supermarket
& $0.065$
& \\
\addlinespace
\hline
\multirow{3}{4.0cm}{Median infectious dose (particles)}
& \multirow{3}{0.8cm}{\centering $d_m$}
& Courtroom
& $300$
& \multirow{3}{*}{\parbox{\linewidth}{
    \cite{vernez_probable_2021,lau_predicting_2022,Prentiss_2022}
}}
\\
& & Care home
& $100$
& \\
& & Supermarket
& $100$
& \\ 
\hline
\multirow{3}{4.0cm}{Breathing rate (m$^3$/s)}
& \multirow{3}{0.8cm}{\centering  $\rho$ }
& Resting
& $1.8 \times 10^{-4}$
& \multirow{3}{\linewidth}{\cite{orton2022comparison,mutsch2022aerosol}}
\\
& & Talking
& $2.2 \times 10^{-4}$
& \\
& & Talking loudly
& $2.5 \times 10^{-4}$
& \\ 
\hline
\multirow{3}{4.0cm}{Aerosol emission rate  (particle/s)}
& \multirow{3}{0.8cm}{\centering  $A$ }
& Resting
& $8$
& \multirow{3}{\linewidth}{\cite{lau_predicting_2022,asadi2019aerosol}}
\\
& & Talking
& $40$
& \\
& & Talking loudly
& $80$
& \\ 
\bottomrule
\end{tabularx}
\end{table}
\section{Results}\label{sec:results}
Here, we compare individual infection risk predictions over time and computation times for VIRIS and VIRIS-3D-L across three single-storey settings. We recompute the VIRIS predictions using the libraries and parameters detailed in Section~\ref{sec:sim_param_process} to ensure a consistent comparison with VIRIS-3D-L.

\subsection{Infection risk comparison}\label{sec:plots_2d_3d_3dl}
\subsubsection{Courtroom}\label{sec:courtroom}
\begin{figure}
    \centering
    \includegraphics[width=0.5\linewidth]{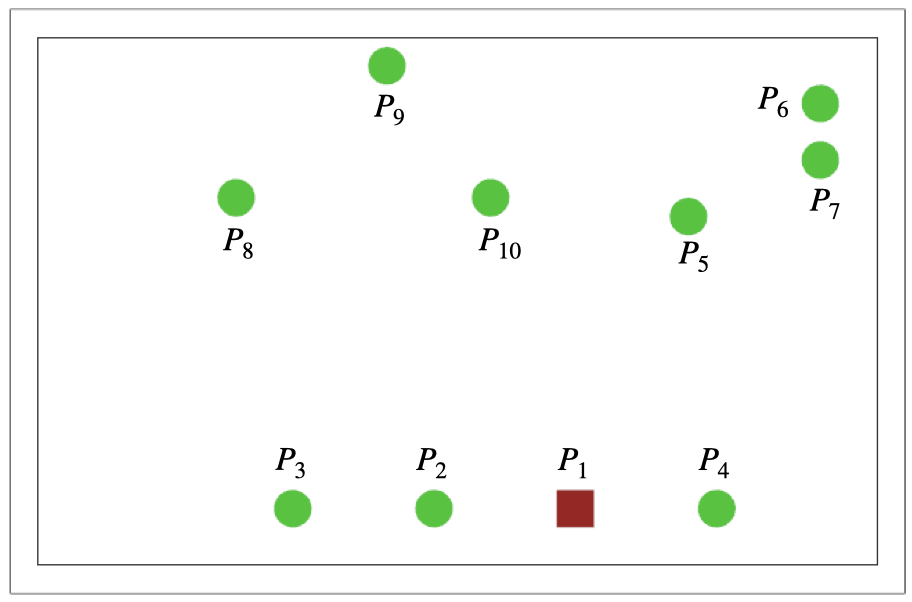}
    \caption{Top view of the courtroom and the location of individuals during the hearing session \cite{vernez_probable_2021}. The infectious individual $P_1$ is shown as a red square, and the susceptible individuals ($P_2$ to $P_{10}$) are shown as green circles.}
    \label{fig:courtroom_OG}
\end{figure}

In this section, as in \cite{xue_viris_2024}, we model the courtroom as a hollow cuboid with no furniture and ten stationary individuals ($P_1$ to $P_{10}$), as shown in Figure \ref{fig:courtroom_OG}. At the start of the simulation, one individual, $P_1$, is infectious (red square in Figure \ref{fig:courtroom_OG}), and nine individuals, $P_2$ to $P_{10}$, are susceptible (green circles in Figure \ref{fig:courtroom_OG}). We run the simulation for the time period 2- 5pm, the duration of the hearing session \cite{vernez_probable_2021}.

\begin{figure}
    \centering
    \includegraphics[width=0.9\linewidth]{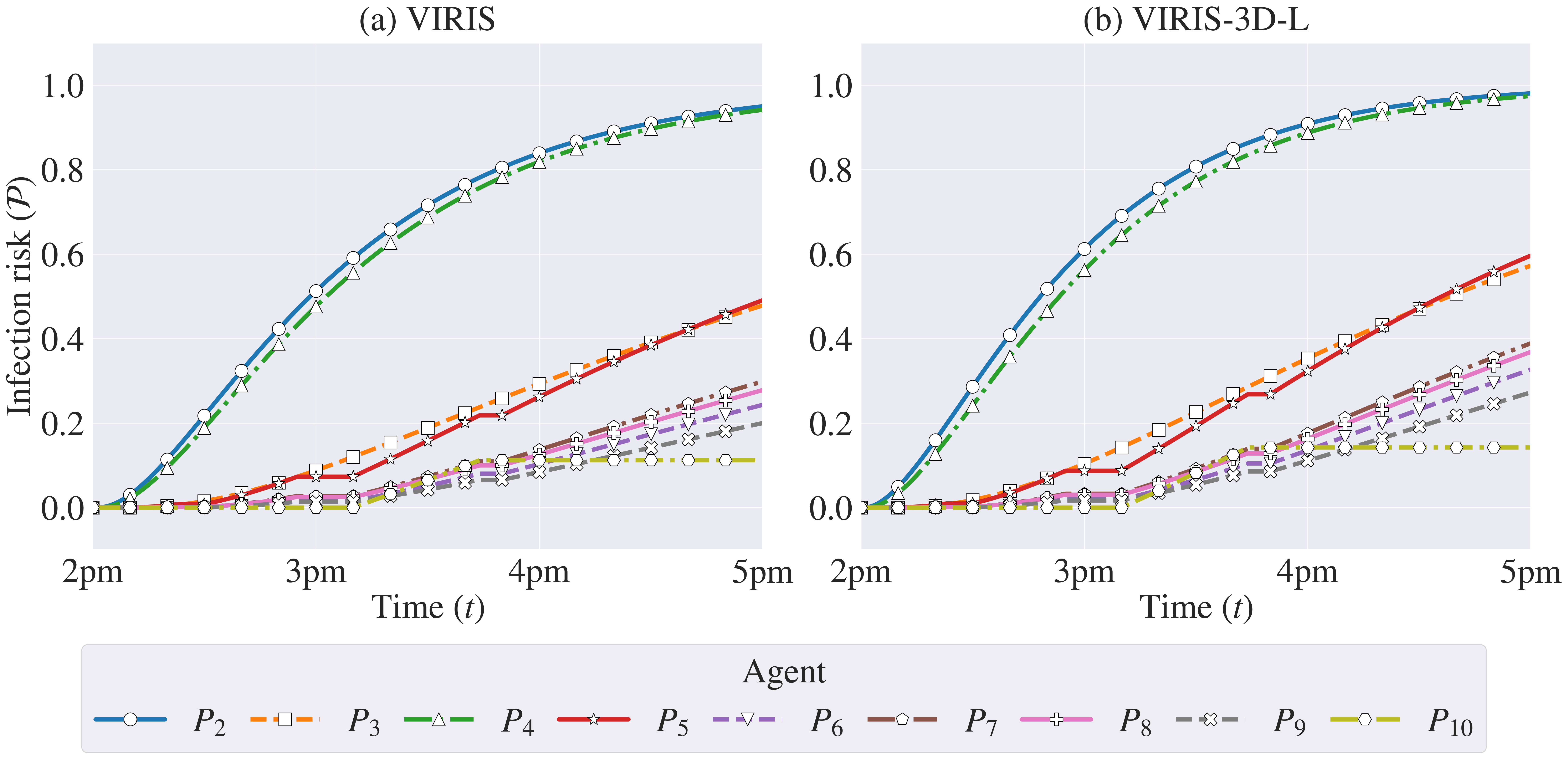}
    \caption{Comparison of the temporal infection risk for nine susceptible individuals ($P_2$ to $P_{10}$) in a courtroom (Figure \ref{fig:courtroom_OG}), estimated using Eq. \eqref{eq:inf_risk_3D} for: a) VIRIS (global removal, 2D domain), and b) VIRIS-3D-L (local removal, 3D domain).}
    \label{fig:courtroom_infection_risk}
\end{figure}

In Figure \ref{fig:courtroom_infection_risk}, we plot the time evolution of infection risk for individuals $P_2$ through $P_{10}$, from 2pm to 5pm, generated using the VIRIS and VIRIS-3D-L frameworks.

We observe that the infection risk evolves similarly over time across the two frameworks, but VIRIS underestimates individuals’ infection risk. For example, at the end of the simulation, the infection risk for $P_3$ and $P_5$ is $48\%$ using VIRIS and $57\%$ using VIRIS-3D-L. Note that the predictions from VIRIS-3D-L better align with the outbreak study \cite{vernez_probable_2021}, which reports that four individuals tested positive for SARS-CoV-2 a few days after the hearing.  Also, infection risk increases more rapidly over time in VIRIS-3D-L than in VIRIS; for example, at 3pm, infection risk for $P_2$ reaches 51\% using VIRIS while it reaches 61\% in VIRIS-3D-L, starting from zero infection risk at 2pm. 

VIRIS-3D-L models airborne transmission more realistically while predicting a higher infection risk and a steeper increase in infection risk than VIRIS; we recommend using VIRIS-3D-L to predict infection risk.

\subsubsection{Care home}\label{sec:carehome}
\begin{figure}
    \centering
    \includegraphics[width=0.5\linewidth]{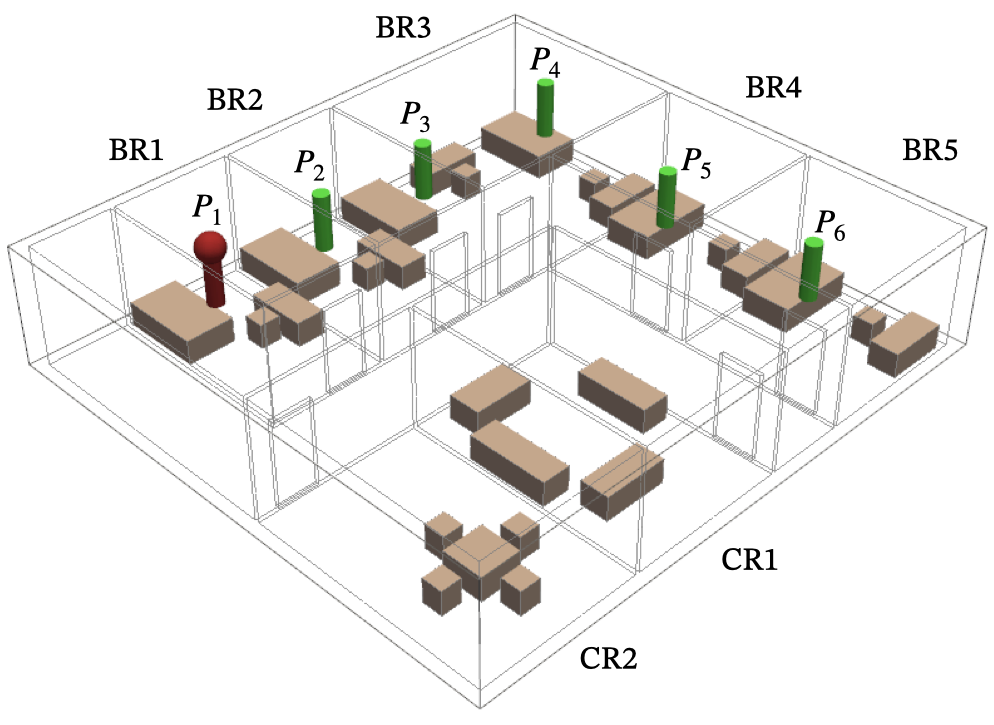}
    \caption{Layout of a small care home with two common rooms (CR1 and CR2) and five bedrooms (BR1-BR5). The infectious resident $P_1$ is shown as a red cylinder with a red spherical head, and susceptible residents $P_2$-$P_6$ are shown as green cylinders.}
    \label{fig:carehome}
\end{figure}

We now study a small care home with five bedrooms (BR1 to BR5), including one double bedroom (BR3), and two common rooms (CR1 and CR2); all rooms are furnished, as shown in Figure \ref{fig:carehome}. The care home has six residents: one is assumed infectious ($P_1$), and five are susceptible ($P_2$ to $P_6$) at the start of the simulation. The simulation runs from 8am to 11pm, covering an entire day of activities. All residents are in one of the common rooms from 9am to 12pm ($P_1,P_3$, and $P_5$ in CR1; $P_2,P_4,$ and $P_6$ in CR2), from 2pm to 5pm ($P_1,P_2$, and $P_3$ in CR1; $P_4,P_5,$ and $P_6$ in CR2), and from 5pm to 8pm ($P_4,P_5,$ and $P_6$ in CR1; $P_1,P_2$, and $P_3$ in CR2). At other times, residents are in their bedrooms.

\begin{figure}
    \centering
    \includegraphics[width=0.9\linewidth]{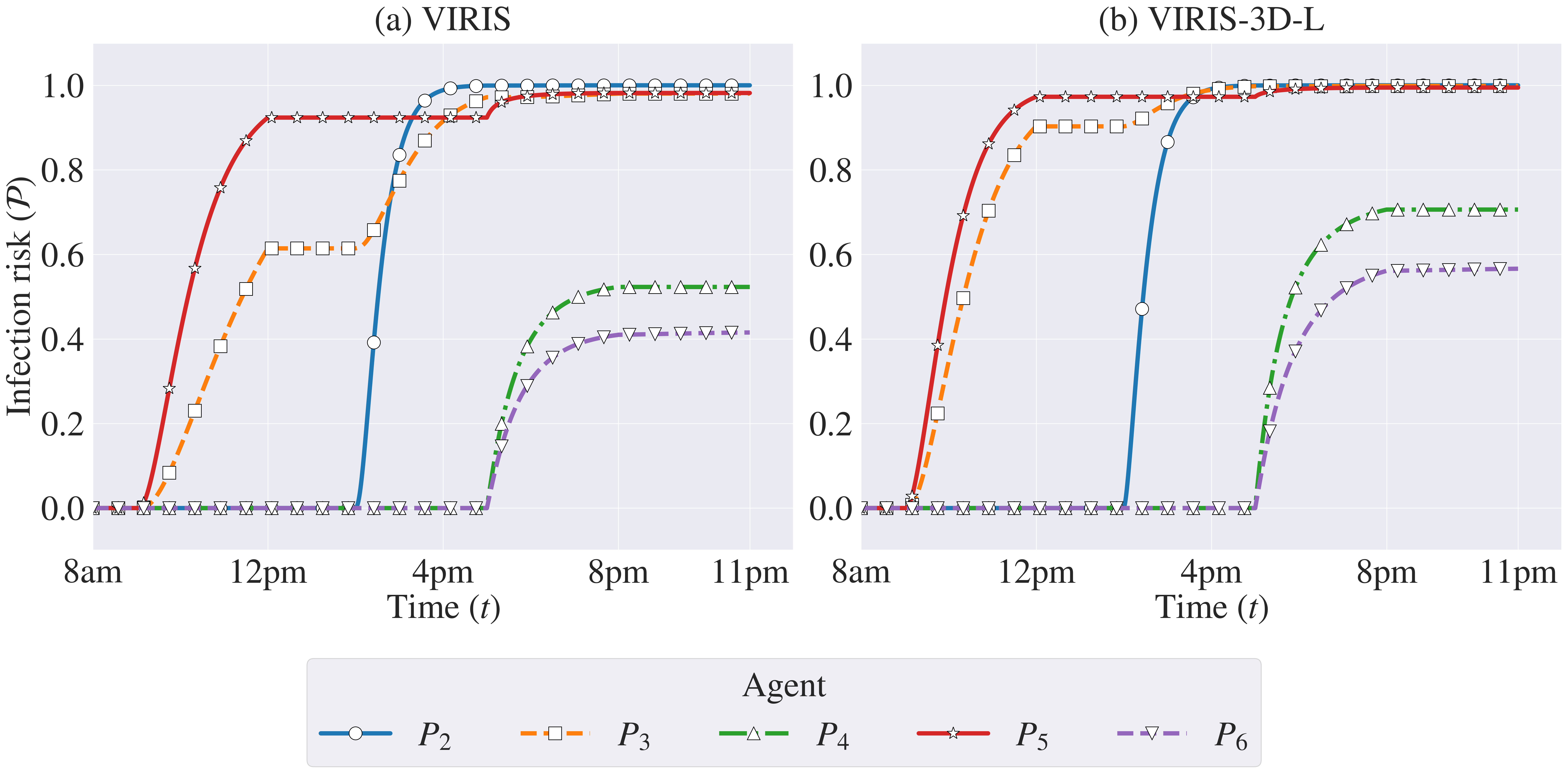}
    \caption{Comparison of the temporal infection risk for five susceptible residents ($P_2$ to $P_{6}$) in a care home (Figure \ref{fig:carehome}), estimated using Eq. \eqref{eq:inf_risk_3D} for: a) VIRIS (global removal, 2D domain), and b) VIRIS-3D-L (local removal, 3D domain).}
    \label{fig:carehome_infection_risk}
\end{figure}

The care home is a more complex environment than the courtroom. The architectural layout (furniture and internal walls) is more complex, and there is some movement of people. Figure \ref{fig:carehome_infection_risk} shows the temporal infection risk for five susceptible residents ($P_2$ to $P_6$) from 8am to 11pm, predicted using the VIRIS and VIRIS-3D-L frameworks, respectively.

At the end of the simulation day, we observe a significant underestimation of infection risk when using VIRIS for $P_4$ and $P_6$ ($52\%$ and $41\%$, respectively) relative to VIRIS-3D-L ($70\%$ and $57\%$, respectively). Note that for the other three residents, infection risk saturates at $100\%$ towards the end of the simulation. However, at earlier times, we see that the infection risk for $P_3$ and $P_5$ is also significantly underestimated by VIRIS. For example, at 12pm, the infection risk of $P_3$ is $61\%$ in VIRIS but $90\%$ in VIRIS-3D-L. When comparing infection risk across all residents at the end of the simulation, we find that VIRIS (78\%) significantly underestimates the average infection risk relative to VIRIS-3D-L (85\%).

Moreover, we observe a rapid increase in infection risk in VIRIS-3D-L compared to VIRIS. For example, for $P_3$, infection risk is 90\% by 12pm with VIRIS-3D-L, whereas with VIRIS it is 61\%. Similarly, for $P_5$, by 12pm infection risk is higher in VIRIS-3D-L (97\%) than in VIRIS (92\%). $P_3$ and $P_5$ are in CR1 with the infectious resident, $P_1$ from 9am to 12pm. During this period, because VIRIS-3D-L predicts higher infectious aerosol concentrations in the breathing zone than VIRIS, infection risk increases faster, and infection risks are higher by 12pm. 

As for the courtroom (Section \ref{sec:courtroom}), VIRIS predicts a significantly lower infection risk and a slower increase in infection risk than VIRIS-3D-L in the care home. Thus, we suggest using VIRIS-3D-L to predict infection risk.

\subsubsection{Supermarket}\label{sec:super_2D_3D}
\begin{figure}
    \centering
    \includegraphics[width=0.5\linewidth]{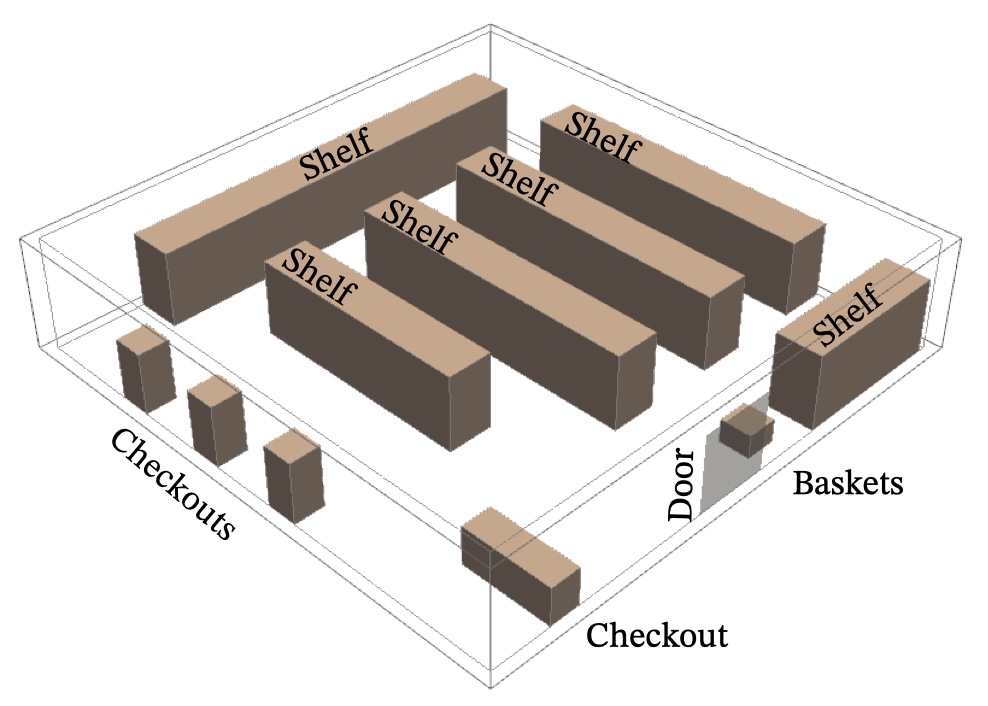}
    \caption{Layout of a small supermarket. It consists of six shelves, four checkout counters, a basket collection point and a door (entry and exit).}
    \label{fig:supermarket}
\end{figure}

Next, we consider a small supermarket (196 $\rm m^2$). As in \cite{xue_viris_2024}, we model the supermarket as a cuboid with six shelves of varying height, four checkout counters, and a basket holder (Figure \ref{fig:supermarket}). The customer flow rate is 60 per hour, and every sixth customer is considered infectious. We simulate people movement over five hours, from 8am to 1pm, but customer entry is restricted until 12pm to ensure customers complete their shopping before the simulation ends. Thus, 240 customers visit the supermarket over four hours, of whom 40 are infectious. Each customer stays in the supermarket for an average of 7.5 minutes, unlike in the care home and the courtroom,  where individuals remain in the domain for the entire simulation.

The supermarket is a high-traffic environment with a more complex architecture than the courtroom and the care home. Also, people movement is more complex and dynamic than in the care home. Customers move continuously, with brief stops to pick up items and pay at the counter, whereas in the care home, residents remain stationary for hours in the bedrooms and common rooms.
\begin{figure}
    \centering
    \includegraphics[width=0.6\linewidth]{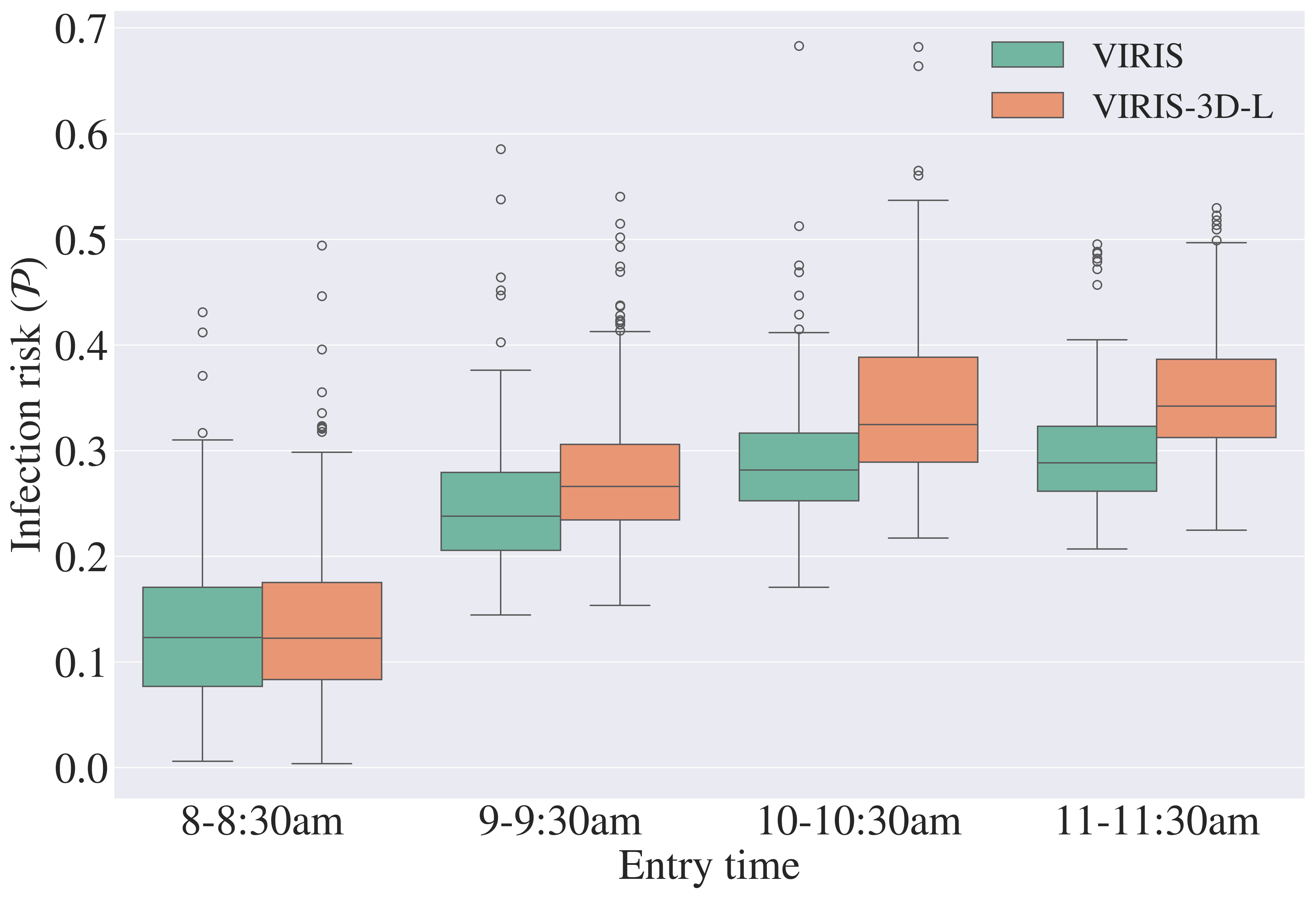}
    \caption{Comparison of customer infection risk at the exit time from the supermarket (Figure \ref{fig:supermarket}), determined using Eq.\eqref{eq:inf_risk_3D} for: a) VIRIS (global removal, 2D domain), and b) VIRIS-3D-L (local removal, 3D domain). The horizontal axis shows four non-overlapping 30-minute entry intervals: 8:00–8:30am, 9:00–9:30am, 10:00–10:30am, and 11:00–11:30am. The horizontal bar in the middle of the box represents the median infection risk. The lower boundary is the 25th percentile ($Q_1$), and the upper boundary is the 75th percentile ($Q_3$) of infection risk for customers who entered during the corresponding entry-time interval. The height of the filled box is the interquartile range, $IQR = Q_3- Q_1$. The hollow circles outside the box are data points beyond $Q_1-1.5IQR$ and $Q_3 + 1.5IQR$, where rare and extreme cases lie.}
    \label{fig:supermarket_infection_risk}
\end{figure}
Here, we present customer temporal infection risks using a different approach. Specifically, we evaluate the infection risk at a customer’s exit time rather than across the entire brief shopping period (7.5 minutes on average). To examine the temporal infection risk, we consider four non-overlapping 30-minute intervals: 8:00–8:30am, 9:00–9:30am, 10:00–10:30am, and 11:00–11:30am, to ensure that customers entering during each interval complete their shopping and exit before the next interval begins. This approach ensures that the entry-time intervals are independent, as customers do not overlap between intervals. Accordingly, Figure \ref{fig:supermarket_infection_risk} shows the distribution of infection risk at the exit for customers whose entry times fall within these four 30-minute intervals, totalling 120 customers.

In both VIRIS and VIRIS-3D-L, infection risk increases over the first two entry time intervals and then saturates. However, we observe a lower infection risk in VIRIS than in VIRIS-3D-L across the four entry-time intervals. The underestimation in median infection risk using VIRIS is smallest for the 8-8:30am interval but increases significantly in later intervals. For example, in the 9-9:30am interval, the median infection risk using VIRIS is 24\%, whereas it is 26\% in VIRIS-3D-L.

Additionally, the upper bounds of the box plots are lower for VIRIS than for VIRIS-3D-L across all entry-time intervals except 8-8:30am, indicating significant underestimation by VIRIS. The corresponding values are 31\%, 39\%, 41\%, and 41\% for VIRIS, compared with 31\%, 41\%, 54\%, and 50\% for VIRIS-3D-L, respectively, from 8-8:30am to 11-11:30am.

As observed in the courtroom (Section \ref{sec:courtroom}) and the care home (Section \ref{sec:carehome}), we also find that the more realistic transmission model in VIRIS-3D-L predicts higher infection risk than VIRIS across all entry-time intervals. These results further support the recommendation to use VIRIS-3D-L for infection risk prediction.

Further, we use a one-sided Wilcoxon \cite{woolson2007wilcoxon} signed-rank test (Appendix \ref{sec:stat_test}) to statistically compare infection risk predictions between the two frameworks. We confirm that VIRIS underestimates infection risk compared with VIRIS-3D-L for all three settings in a statistically significant way.

To assess the effect of extending the model from two to three dimensions while retaining global surface deposition, we compare VIRIS with the intermediate VIRIS-3D framework across the three indoor settings. The average infection risks across individuals at the end of the simulations are 44\%, 78\%, and 25\% for VIRIS, compared with 45\%, 79\%, and 26\% for VIRIS-3D, in the courtroom, care home, and supermarket, respectively. The predicted infection risk also increases slightly more rapidly over time in VIRIS-3D than in VIRIS, although the differences between the two models remain small. Overall, these results suggest that extending the model to three dimensions, while retaining global surface deposition, has only a modest effect on predicted infection risk.

\subsection{Computation time comparison}\label{sec:comp_time}
In Table \ref{tab:comp_time}, we compare the computational times for VIRIS and VIRIS-3D-L for the three settings. As expected, the computation time for VIRIS-3D-L is higher than that for VIRIS. This is because solving the reaction-diffusion equation \eqref{eq:rd_viris_2d} takes longer in a 3D domain than in a 2D domain (Table \ref{tab:process_comp_time_combined_compact} in Appendix \ref{app:comp_time}). Specifically, total computation time increases by 6.4 minutes for the courtroom, 19.8 minutes for the care home, and an hour for the supermarket. The care home simulation involved 15 hours of moving residents, compared with three hours of static individuals in the courtroom. The increase in computation time for the supermarket is highest, reflecting its dynamic environment, 240 customers, and complex layout, compared with the care home’s six moderately dynamic residents and ten static individuals in the courtroom.
\begin{table}[ht]
    \small
    \centering
    \begin{tabular}{lcc}
        \toprule
        & \multicolumn{2}{c}{\textbf{Computation time (min)}} \\
        \cmidrule(lr){2-3}
        \textbf{Setting}
        & \textbf{VIRIS}
        & \textbf{VIRIS-3D-L} \\
        \midrule
        Courtroom   & 1.6  & 8.0\\
        Care home   & 9.2& 29.0\\
        Supermarket & 19.5& 80.6\\
        \bottomrule
    \end{tabular}
    \caption{Computation time in minutes for three single-storey settings (courtroom, care home, supermarket) using the VIRIS and VIRIS-3D-L modelling frameworks.}
    \label{tab:comp_time}
\end{table} 
\section{Conclusion} \label{sec:conclusion}
In this paper, we extended the airborne infection risk modelling framework VIRIS \cite{xue_viris_2024} to the VIRIS-3D-L framework by modifying two assumptions to make them more realistic. First, we solved the reaction-diffusion equation modelling airborne transmission \eqref{eq:rd_viris_3d} in a 3D domain instead of a 2D domain, and second, we removed the global removal, first-order, surface-deposition sink term from this equation and included instead a Robin boundary condition that captures local aerosol removal at surfaces. We compared individuals' infection risk predictions over time and computation times for VIRIS and VIRIS-3D-L, in three single-storey settings of increasing complexity in architectural layout and people movement: a courtroom, a care home, and a supermarket. 

For all three settings, VIRIS significantly underestimated the average infection risk across individuals compared with VIRIS-3D-L. For the courtroom, the infection risk was 44\% versus 51\%, for the care home 78\% versus  85\% and for the supermarket 25\% versus 28\%, respectively, for VIRIS and VIRIS-3D-L. Moreover, in all settings, the infection risk rises more slowly over time in VIRIS compared to VIRIS-3D-L. These differences arise because VIRIS assumes uniform aerosol concentrations over the building height and applies surface-deposition removal throughout the domain, including the breathing zone, which lowers the predicted infectious aerosol concentrations. By resolving vertical concentration gradients and correctly delegating removal only to surfaces, VIRIS-3D-L predicts higher breathing zone concentrations and greater spatial variation in infection risk, particularly in the supermarket, where furniture heights vary. In VIRIS-3D-L, computation time increases by approximately fourfold on average, but it is still very doable on a standard laptop, with the longest simulation taking just under 1.5 hours.

The lower infection risk estimates in VIRIS overestimate safe exposure durations and would lead to less effective interventions for epidemics. Thus, VIRIS-3D-L is recommended for infection risk prediction and intervention assessment, particularly in settings with varying heights of people and furniture or frequent people movement.

Although only single-storey settings are considered here, VIRIS-3D-L can be applied to multi-storey settings. Future work could validate VIRIS-3D-L against measured aerosol concentrations and outbreak data, investigate airflow-driven advection in well-ventilated spaces, and quantify how local and global representations of ventilation removal affect infection risk.\\

\small \noindent \textbf{Funding statement. }This work is supported by the Natural Environment, Biotechnology and Biological Sciences and Medical Research councils (NERC, BBSRC and MRC) [grant number: NE/X016714/1] as part of the One Health for One Environment: an A-Z Approach for Tackling Zoonoses (`OneZoo') Centre for Doctoral Training. This work is also supported by an EPSRC Impact Acceleration Account grant, under grant number EP/X525522/1.\\

\small \noindent \textbf{Competing interests. }The authors report no conflict of interest.
\bibliographystyle{plain}
\bibliography{references}

\appendix

\section{Layout and people schedules for three single-storey settings}\label{layouts_3}
In this section, we describe the layout and schedule of people in three single-storey settings, used in this study for comparing VIRIS and VIRIS-3D-L frameworks. 

\subsection{Courtroom}\label{court_layout}
We consider a single-storey courtroom in the Canton of Vaud, Switzerland. In \cite{vernez_probable_2021}, extensive data from a hearing held from 2pm to 5pm on 30th October 2020 were analysed to study airborne transmission of SARS-CoV-2. The hearing was attended by ten individuals, and within a week of the hearing, four individuals tested positive. The dimensions of the courtroom are $ 8. 9 \text{ m }(l) \times 5.6  \text{ m }(w) \times 3 \text{ m } (h)$. We assume there are no openings in the windows or doors, as they were closed during the hearing for confidentiality. We label the individuals as $P_1,P_2, P_3, P_4, P_5, P_6, P_7, P_8, P_9,$ and $ P_{10}$. All individuals were assigned seats in the courtroom, at least $ 1. 5 $ m apart from each other, as shown in Figure \ref{fig:courtroom_OG}, and they remained seated throughout the hearing. $ P_1 $ was the only infectious individual, represented by a red square in Figure \ref{fig:courtroom_OG}, and is considered a superspreader \cite{ vernez_probable_2021,lau_predicting_2022}. Within a week of the hearing, $ P_2, P_3, P_4 $, and $ P_5 $ tested positive. However, $P_5$ had contact with another positive case before the hearing. The individuals $P_1$ to $P_9$ were present in the courtroom for the entire hearing, whereas $P_{10}$, the witness, was present only for the last $34$ minutes of the hearing. During the hearing, there were four breaks: 2pm to 2:05pm, 2:23pm to 2:30pm, 2:55pm to 3:10pm, and 3:44pm to 3:50pm. During the breaks, $P_1$ to $P_4$ remained in the courtroom, and others went into the waiting hall outside the courtroom. None of the individuals wore a mask during the hearing period. Additionally, the mechanical ventilation system was broken during the hearing, leading to poor ventilation in the courtroom. 

\subsection{Care home}\label{carehome_layout}
The second single-storey setting is a small care home with dynamic residents. The care home consists of five bedrooms and two common rooms connected by a corridor, with six residents, as shown in Figure \ref{fig:carehome}. The overall dimensions of the care home are $14\text{ m } (l) \times 13\text{ m } (w) \times 3\text{ m } (h)$. Two residents share one of the five bedrooms, and the remaining four bedrooms are occupied by a single resident, as shown in Figure \ref{fig:carehome}. We simulate an entire day from 8am to 11pm in the care home. The residents are labelled $P_1, P_2, P_3, P_4, P_5,$ and $P_6$, and their assigned bedrooms are shown in Figure \ref{fig:carehome}. We assume that $P_1$ is infectious at the start of the simulation. Before 9am, all residents are in their respective bedrooms. From 9am to 12pm, $P_1, P_3,$ and $P_5$ are in the common room labelled CR1, and $P_2, P_4,$ and $P_6$ are in the other common room labelled CR2 (Figure \ref{fig:carehome}). After 12pm, all residents return to their respective bedrooms. From 2pm to 5pm, $P_1, P_2,$ and $P_3$ go to CR1, and $P_4, P_5,$ and $P_6$ go to CR2. Then the residents exchange common rooms; $P_1, P_2,$ and $P_3$ go to CR2, and $P_4, P_5,$ and $P_6$ go to CR1, and they stay there from 5pm to 8pm. After that, all residents retire to their respective bedrooms. We assume that each resident is talking when they are in the common rooms and resting when they are in the bedroom.

\subsection{Supermarket}\label{super_layout}
The third single-storey setting is a small supermarket, characterised by a highly dynamic environment and a complex indoor layout. The supermarket differs from the courtroom and care home because individuals do not remain indoors for the entire simulation duration and are continuously walking. The overall dimensions of the supermarket are $14\text{ m }(l) \times 12\text{ m } (w) \times 3\text{ m } (h)$, as shown in Figure \ref{fig:supermarket}. The simulation period is five hours, from 8am to 1pm. We assume that 60 customers enter the supermarket per hour, i.e., one customer every minute \cite{rybaczewska2021covid,Lumina2025ConvenienceReport}. We restrict entry to the supermarket until 12pm, so total footfall is 240 customers.

A customer's trajectory begins when they enter through the door. Then the customer goes to a nearby basket holder to collect a basket, which takes $15$ seconds. After that, the customer moves around the supermarket to pick exactly six items from the shelves. The time to pick an item is sampled from a uniform distribution $U(40,80)$. For each customer, six items are chosen at random from the available choices. After collecting the items, the customer goes to the checkout counter. There are four counters. Depending on availability, one counter is randomly chosen. The checkout time is also sampled from a uniform distribution $U(40,80)$. Finally, the customer leaves the supermarket via the door. The entry and exit doors are the same. Each customer spends 7.5 minutes on average in the supermarket. We assume that one in every 6 customers entering the supermarket is infectious, i.e., 40 infectious customers during the simulation period \cite{ONS2022CovidSurvey}. So, at the hourly level, ten infectious customers enter the supermarket.

\section{Courtroom outbreak data - Model calibration}\label{app:courtroom_mu_dm}
\begin{figure}
    \centering
\includegraphics[width=0.9\linewidth]{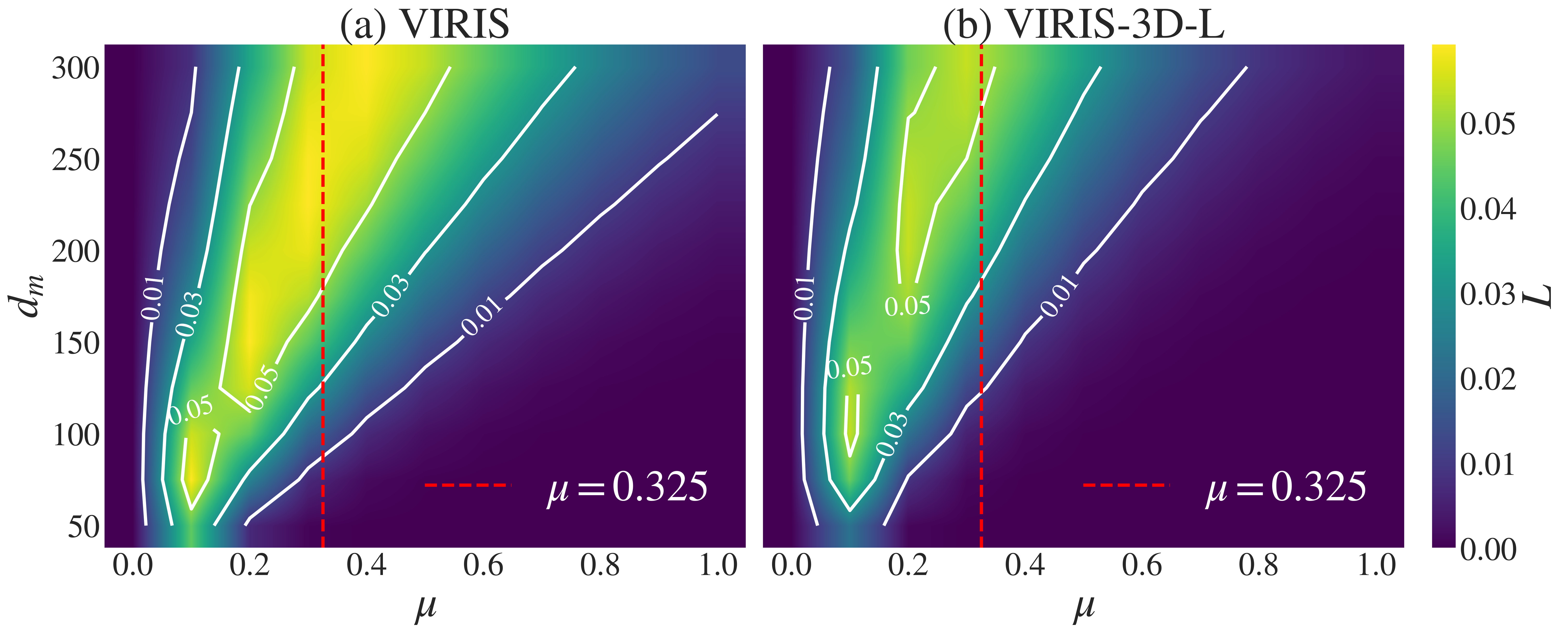}
    \caption{Comparison of likelihood values using Eq.~\eqref {eq:likelihood} to reproduce the courtroom outbreak data \cite{vernez_probable_2021} for the parameters $\mu$ and $d_m$ in VIRIS and VIRIS-3D-L.}
    \label{fig:mu_dm_comparison}
\end{figure}

In this section, we conduct a parameter sweep over $\mu$, the fraction of virus-laden aerosols in the breath of an infected individual, and $d_m$, the dose needed to infect 50\% of susceptible individuals under specific conditions. We evaluate how well each parameter combination reproduces the courtroom outbreak data \cite{vernez_probable_2021}, in which individuals $P_2$ through $P_5$ became infected after the hearing session, while individuals $P_6$ through $P_{10}$ remained uninfected. We calculate the likelihood for each parameter combination as follows:
\begin{align}\label{eq:likelihood}
L = \prod_{i=2}^{5} \mathcal{P}i \prod_{i=6}^{10} (1-\mathcal{P}_i).
\end{align}
Figure \ref{fig:mu_dm_comparison} shows the likelihood values for VIRIS and VIRIS-3D-L. Although we sweep over both parameters, we use $\mu=0.325$, the assumed superspreader value reported in the outbreak data and indicated by the red dashed line, to identify $d_m$. At this fixed value of $\mu$, the likelihood is highest at $d_m=300$ for both models. We therefore use $\mu=0.325$ and $d_m=300$ for the courtroom simulations.

\section{Concentration of infectious aerosols}
\begin{figure}[htbp]
    \centering
    \begin{tabular}{cc}
        \multicolumn{2}{c}{VIRIS and VIRIS-3D-L} \\[0.4em]

        \multicolumn{2}{c}{Courtroom} \\[0.2em]

        \begin{subfigure}[t]{0.35\textwidth}
            \centering
            \includegraphics[width=\textwidth]{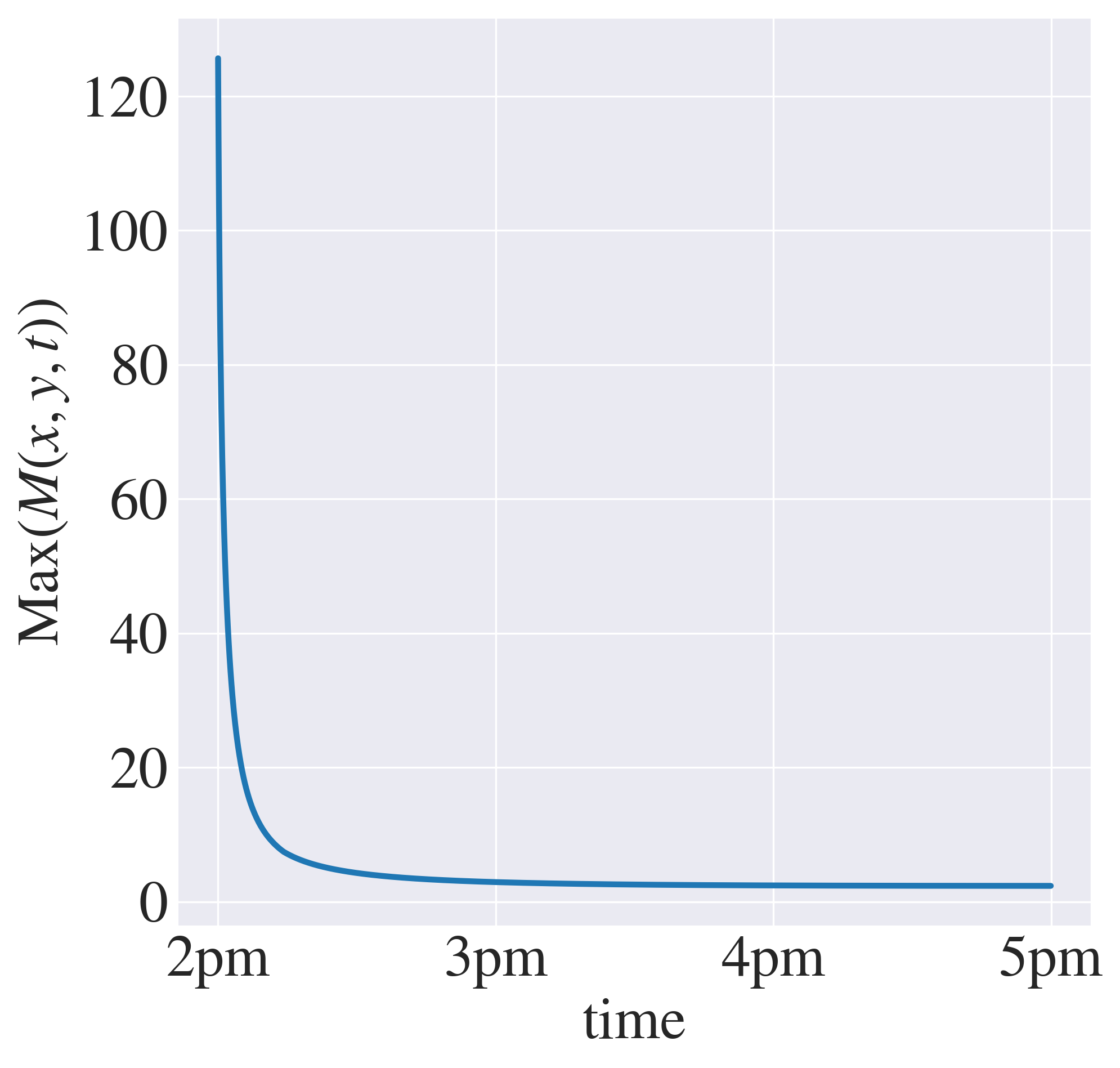}
            \caption{Maximum}
            \label{fig:virus_conc_max_court}
        \end{subfigure}
        &
        \begin{subfigure}[t]{0.35\textwidth}
            \centering
            \includegraphics[width=\textwidth]{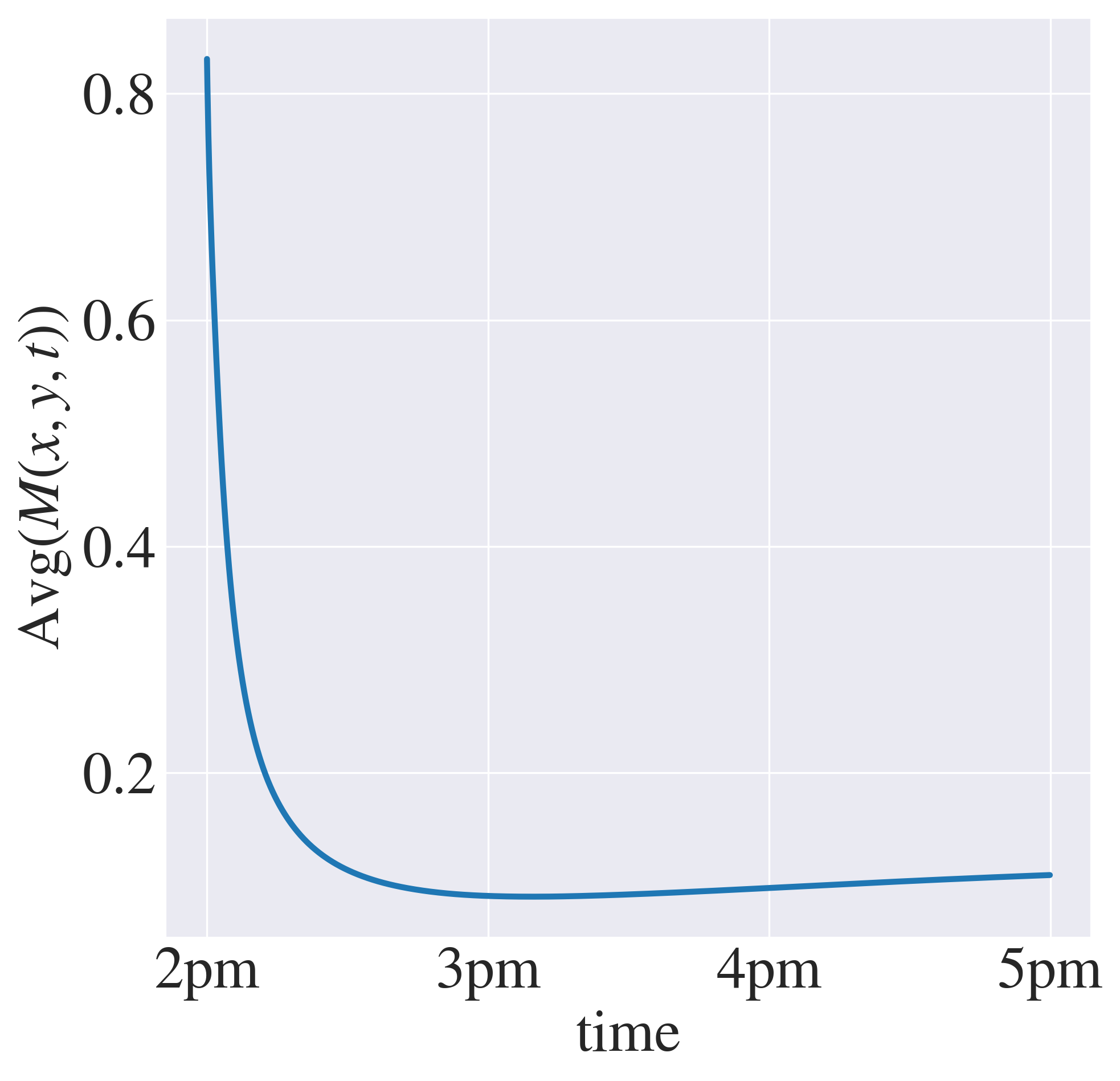}
            \caption{Average}
            \label{fig:virus_conc_mean_court}
        \end{subfigure}
        \\[0.8em]

        \multicolumn{2}{c}{Care home} \\[0.2em]

        \begin{subfigure}[t]{0.35\textwidth}
            \centering
            \includegraphics[width=\textwidth]{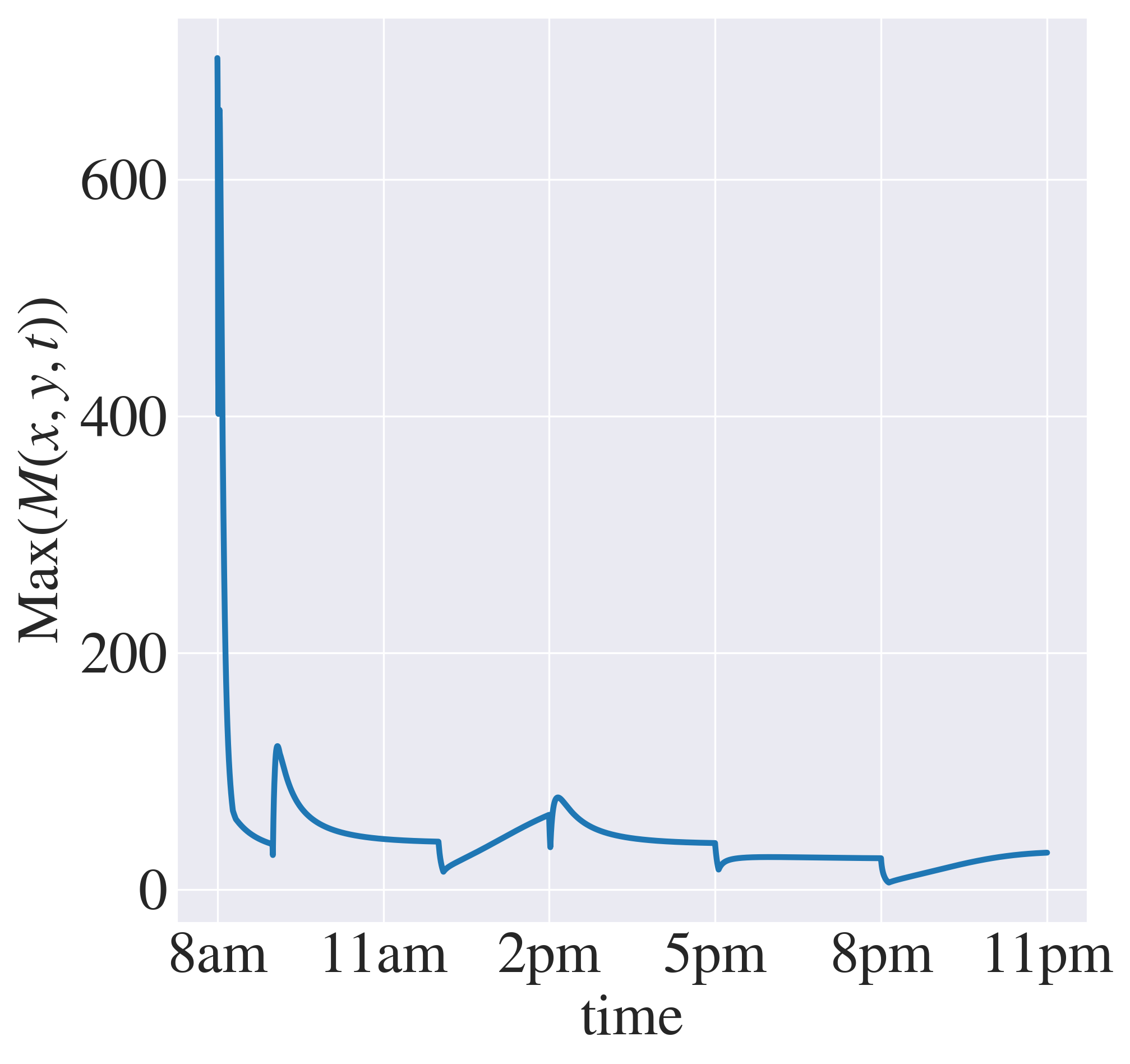}
            \caption{Maximum}
            \label{fig:virus_conc_max_care}
        \end{subfigure}
        &
        \begin{subfigure}[t]{0.35\textwidth}
            \centering
            \includegraphics[width=\textwidth]{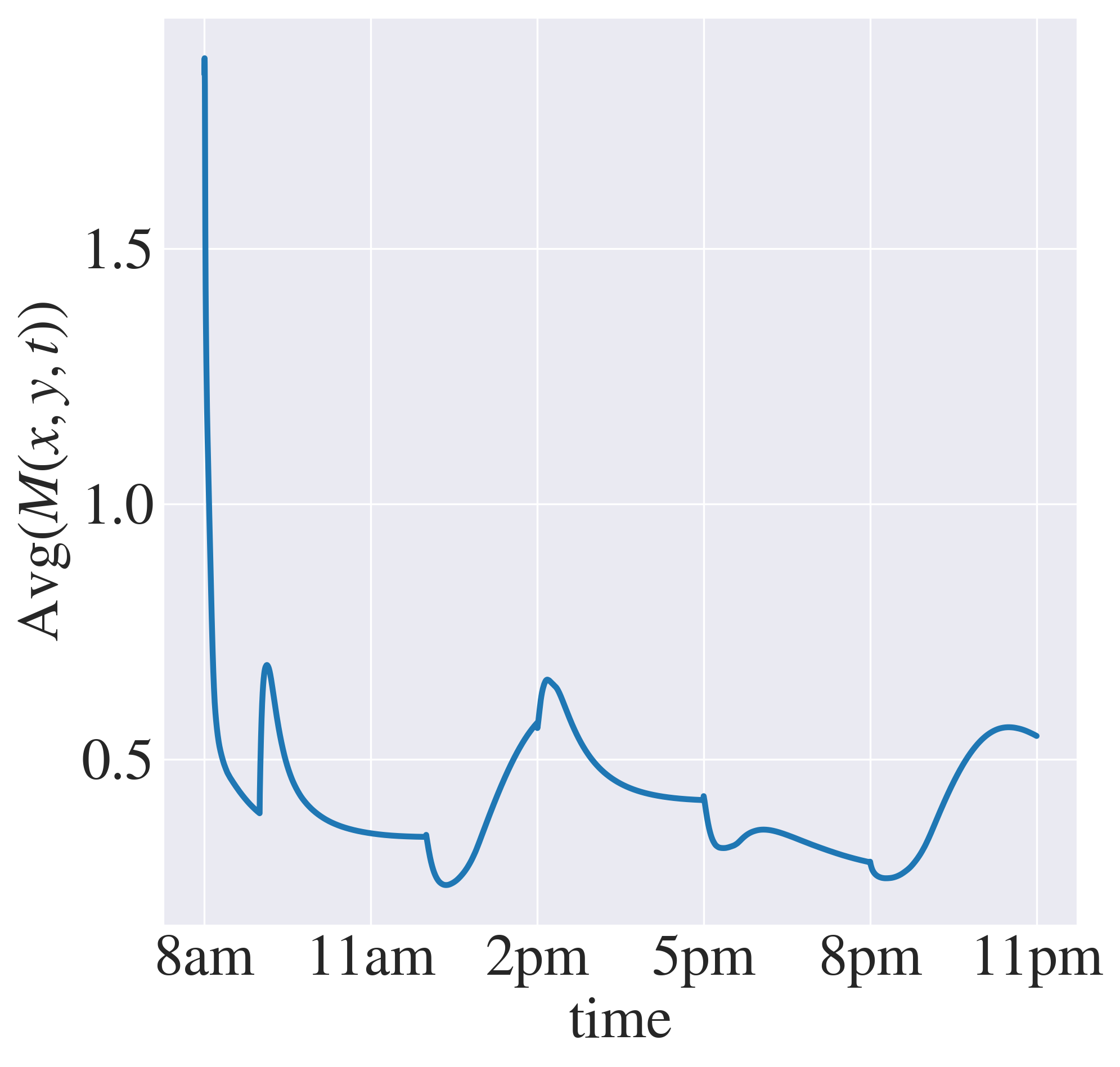}
            \caption{Average}
            \label{fig:virus_conc_mean_care}
        \end{subfigure}
        \\[0.8em]

        \multicolumn{2}{c}{Supermarket} \\[0.2em]

        \begin{subfigure}[t]{0.35\textwidth}
            \centering
            \includegraphics[width=\textwidth]{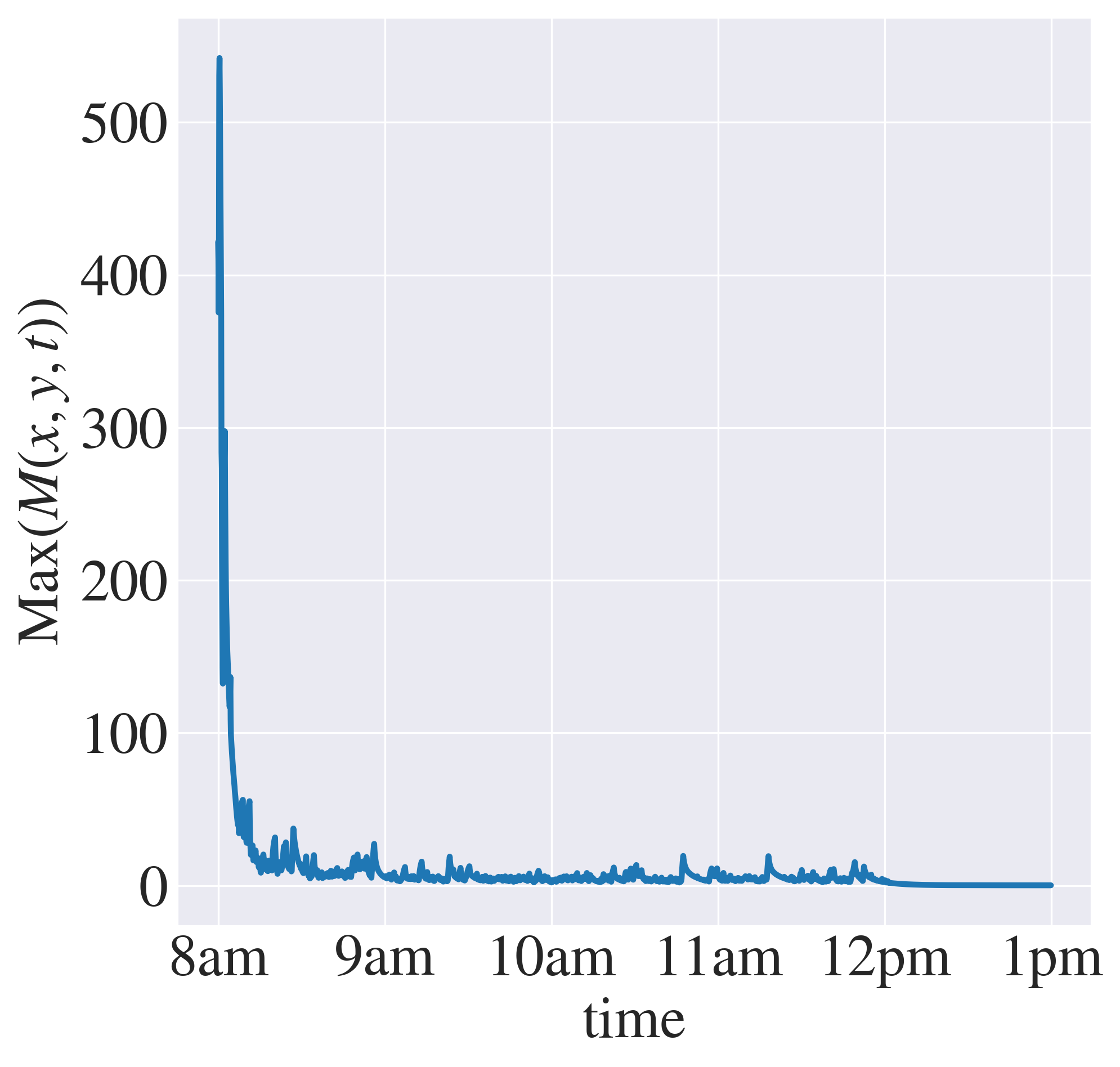}
            \caption{Maximum}
            \label{fig:virus_conc_max_super}
        \end{subfigure}
        &
        \begin{subfigure}[t]{0.35\textwidth}
            \centering
            \includegraphics[width=\textwidth]{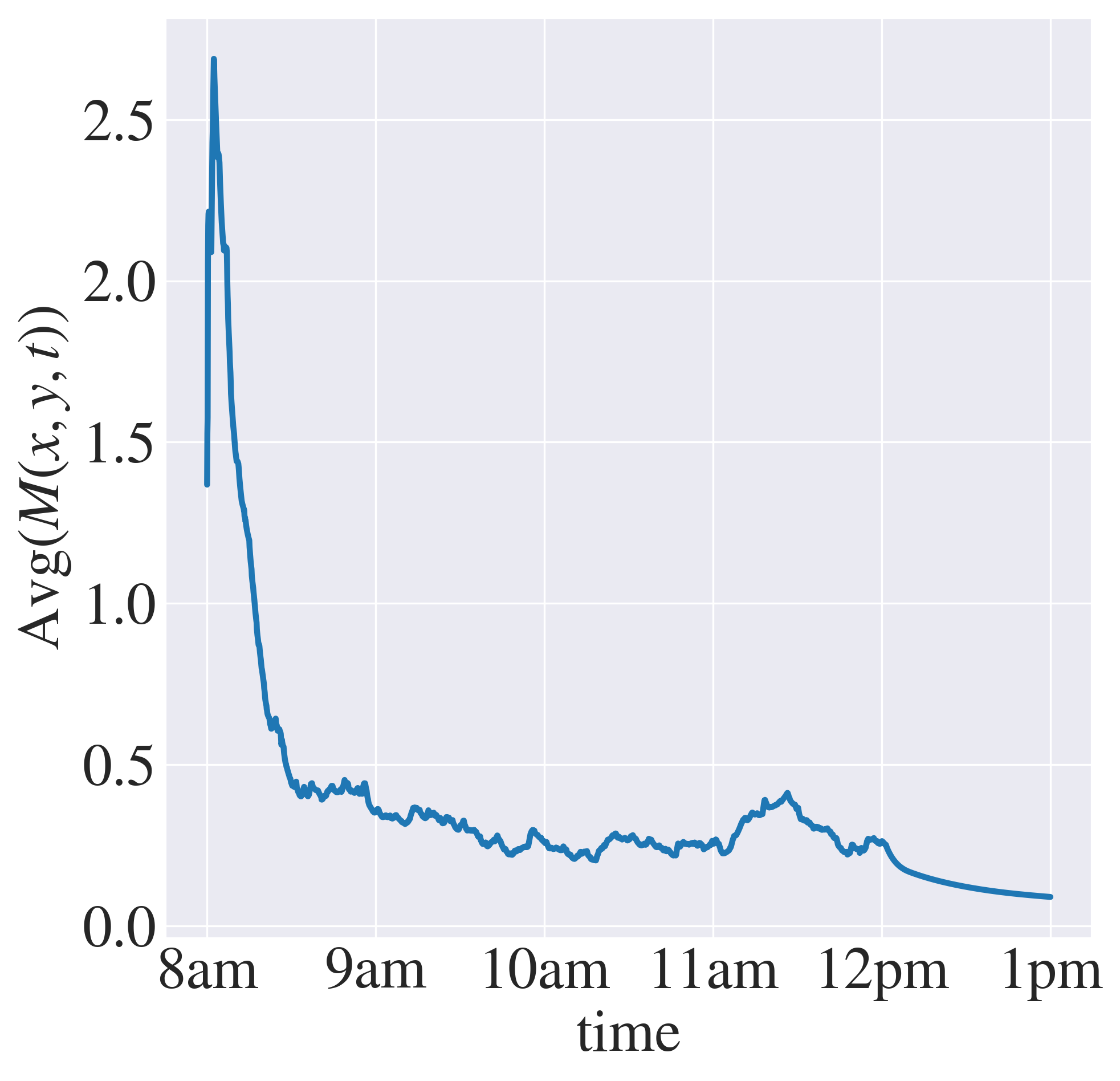}
            \caption{Average}
            \label{fig:virus_conc_mean_super}
        \end{subfigure}
        \\
    \end{tabular}

    \caption{Temporal variation in the maximum and average absolute differences in virus concentration between VIRIS and VIRIS-3D-L. The first row is for the courtroom, the second for the care home, and the third for the supermarket.}
    \label{fig:virus_conc_comparison_combined}
\end{figure}

In this section, we compare the temporal evolution of the infectious aerosol concentration across the three single-storey settings (courtroom, care home, and supermarket) by solving Eq. \eqref{eq:rd_viris_2d} for VIRIS and Eq. \eqref{eq:rd_viris_3d} for VIRIS-3D-L. We estimate the absolute differences in spatio-temporal concentration between VIRIS and VIRIS-3D-L, as follows:
\begin{enumerate}
    \item We scale the spatial concentration values at each time point for each of the two modelling frameworks by dividing by the total concentration at that time point. For VIRIS, with $\Omega \in \mathrm{R}^2$, the scaled concentration values are given by,
    \begin{align*}
    C_{\text{scaled}} (x,y,t) &= \frac{C (x,y,t)}{\int_{\Omega}C (x,y,t) dS}.
    \end{align*}
    For VIRIS-3D-L, with $\Omega \in \mathrm{R}^3$, 
    \begin{align*}
    L_{\text{scaled}} (x,y,z,t) &= \frac{C (x,y,z,t)}{\int_{\Omega}C (x,y,z,t) dV}.
    \end{align*}
    \item For the VIRIS and VIRIS-3D-L comparison, we take the average of $L_{\text{scaled}}(x,y,z,t)$ over the height to bring it to the 2D domain of VIRIS, as follows,
    \begin{align*}
    \bar{L}_{\text{scaled}}(x,y,t) = \int_{0}^{3} L_{\text{scaled}} (x,y,z,t) dz.
    \end{align*}
    \item Then, we estimate the absolute difference between scaled spatio-temporal concentration values.  The absolute difference is given by,
    \begin{align}
    \text{VIRIS and VIRIS-3D-L: } M(x,y,t) &= |\bar{L}_{\text{scaled}}(x,y,t) - C_{\text{scaled}} (x,y,t) |. \label{eq:2d_3dl}
    \end{align}
\end{enumerate}
In Figure \ref{fig:virus_conc_comparison_combined}, we plot the average and maximum values of the absolute differences, $M(x,y,t)$, estimated using equation \eqref{eq:2d_3dl} to compare the two frameworks across the three settings. We find that the concentration difference is positive and decreases with time in all three settings, indicating that VIRIS underestimates compared to VIRIS-3D-L. For the care home, we observe jumps, corresponding to changes in location of residents from bedroom to common room or vice versa. Similarly, we observe small jumps in the supermarket, corresponding to customers entering and exiting.

\section{Statistical comparison of infection risk}\label{sec:stat_test}
In this section, we statistically compare individual-level temporal infection-risk estimates from the VIRIS and VIRIS-3D-L models for three single-storey settings. We evaluate infection risk using Eq.~\eqref{eq:inf_risk_3D}; the resulting values are bounded between $0$ and $1$ and are not normally distributed. Thus, we perform a non-parametric test for paired observations, the Wilcoxon signed-rank test \cite{woolson2007wilcoxon}, to compare individual-level temporal infection risk estimates.

For each individual $i$ and time $t \in [0,T]$, where $T$ is the simulation end time for the corresponding setting, we define the paired difference in infection risk for two model comparisons as follows:
\begin{align}
    \text{VIRIS versus VIRIS-3D-L:} \quad
    & d_{\text{2D-3DL}}(i,t)
      = p_{\text{VIRIS}}(i,t) - p_{\text{VIRIS-3D-L}}(i,t).
\end{align}
We test the hypotheses at the significance level $\alpha = 0.05$. We first assess whether the two models in each pair produce systematically different infection-risk estimates using a two-sided Wilcoxon signed-rank test.
\begin{hypothesis}
\label{hyp:two-sided}
\null\par
\noindent \textbf{Infection risks are different for the two models.} \par
\noindent $H_0$:  $\operatorname{median}\!\left(d_{\text{2D-3DL}}(i,t)\right)=0$. \par
\noindent $H_1$: $\operatorname{median}\!\left(d_{\text{2D-3DL}}(i,t)\right)\neq 0$.
\end{hypothesis}
The test statistics and $p$-values are presented in Table~\ref{tab:wilcoxon_results_3d_3dl} for the VIRIS-3D-L versus VIRIS comparison. For all three settings, the two-sided $p$-values are below $0.05$. Therefore, we reject $H_0$ and conclude that the paired infection-risk estimates differ between VIRIS and VIRIS-3D-L.

To determine the direction of the difference, we then perform one-sided Wilcoxon signed-rank tests. Negative values of $d_{\text{2D-3DL}}(i,t)$ indicate that VIRIS predicts lower infection risk estimates than VIRIS-3D-L, and vice versa.
\begin{hypothesis}
\label{hyp:one-sided-over}
\null\par
\noindent \textbf{VIRIS overestimates.} \par
\noindent $H_0$: $\operatorname{median}\!\left(d_{\text{2D-3DL}}(i,t)\right)\leq 0$. \par
\noindent $H_1$: $\operatorname{median}\!\left(d_{\text{2D-3DL}}(i,t)\right)>0$. \par
\noindent Equivalently, rejecting $H_0$ implies that VIRIS systematically overestimates infection risk relative to VIRIS-3D-L.
\end{hypothesis}
\begin{hypothesis}
\label{hyp:one-sided-under}
\null\par
\noindent \textbf{VIRIS underestimates.} \par
\noindent $H_0$: $\operatorname{median}\!\left(d_{\text{2D-3DL}}(i,t)\right)\geq 0$. \par
\noindent $H_1$: $\operatorname{median}\!\left(d_{\text{2D-3DL}}(i,t)\right)<0$. \par
\noindent Equivalently, rejecting $H_0$ implies that VIRIS systematically underestimates infection risk relative to VIRIS-3D-L.
\end{hypothesis}
The one-sided test results are presented in Table~\ref{tab:wilcoxon_results_3d_3dl}. For all three settings, the $p$-values for the overestimation hypothesis \eqref{hyp:one-sided-over} are $1$, whereas the corresponding $p$-values for the underestimation hypothesis \eqref{hyp:one-sided-under} are below $0.05$. Thus, VIRIS significantly underestimates infection risk relative to VIRIS-3D-L for all three settings.
\begin{table}[htbp]
\centering
\small
\begin{tabular}{lrrrrr}
\hline
\textbf{Setting} &
\textbf{Statistic} &
\textbf{Two-sided $p$} &
\textbf{Overest. $p$} &
\textbf{Underest. $p$} &
\textbf{Median diff.} \\
\hline
Courtroom &
$7.7 \times 10^3$ &
$0.0$ &
$1.0$ &
$0.0$ &
$-0.03$ \\

Care home &
$3.4 \times 10^6$ &
$0.0$ &
$1.0$ &
$0.0$ &
$-0.01$ \\

Supermarket &
$1.8 \times 10^{10}$ &
$0.0$ &
$1.0$ &
$0.0$ &
$0.0$ \\
\hline
\end{tabular}
\caption{Wilcoxon signed-rank test results for three single-storey settings to compare VIRIS and VIRIS-3D-L.}
\label{tab:wilcoxon_results_3d_3dl}
\end{table}

\section{Computation time}\label{app:comp_time}
In this section, we compare the computation time for each component of the two modelling frameworks, VIRIS and VIRIS-3D-L, for three single-storey settings. The difference in computation time (in minutes) between VIRIS-3D-L and VIRIS-3D is provided in Table \ref{tab:process_comp_time_combined_compact}.
\begin{table}[ht]
    \centering
    \small
    \begin{tabular}{cccc}
        \toprule
        \textbf{Process}
        & \textbf{Courtroom}
        & \textbf{Care home}
        & \textbf{Supermarket} \\
        \midrule
        Topology creation
        & $-7.8 \times 10^{-4}$&$1.0 \times 10^{-2}$&$1.0 \times 10^{-1}$\\
         Schedule creation
        & $-4.7 \times 10^{-2}$& $3.7 \times 10^{-2}$& $2.4 \times 10^{-2}$\\
         Mesh generation
        &$2.9 \times 10^{-5}$&$0.7\times 10^{-5}$& $5.0 \times 10^{-6}$\\
         Individuals' trajectory generation
        & $-3.7 \times 10^{-2}$& $-2.1 \times 10^{-1}$&$-4.3 \times 10^{-2}$\\
         Reaction-diffusion solution
        &$6.4$& $20.0$&$61.1$\\
         Infection risk estimation
        & $7.8 \times 10^{-3}$&$-4.1 \times 10^{-2}$&	$5.6 \times 10^{-3}$\\
        \bottomrule
    \end{tabular}

    \caption{Difference in computation time, in minutes, between VIRIS-3D-L and VIRIS frameworks, for each processing step for three single-storey settings.}
    \label{tab:process_comp_time_combined_compact}
\end{table}
\end{document}